\documentclass[11pt,draftcls]{IEEEtran}
\onecolumn
\usepackage[mathcal]{euscript}
\usepackage{bbm,dsfont,mathrsfs,fixmath}
\usepackage{stmaryrd}
\usepackage{amsmath,amstext,amsfonts,amssymb}
\usepackage{epsfig,float}
\usepackage{graphics}
\usepackage{epstopdf}
\usepackage{cite}
\usepackage{psfrag}
\usepackage{subfigure}
\usepackage{url}
\usepackage{shadow,color,pifont,times,rotate}

\newcommand{\mb}{\mathbf}

\newcommand{\mc}{\mathcal}

\def\ex{\times }

\def\esp{{\mathbb E}}  
 \DeclareMathAlphabet{\mathpzc}{OT1}{pzc}{m}{it}
\newtheorem{theorem}{Theorem}[section]

\newtheorem{conjecture}{Conjecture}
\newtheorem{corollary}{Corollary}
\newtheorem{remark}{Remark}
\newtheorem{definition}{Definition}

\usepackage{enumerate}

\begin{document}
  \footernote{To appear in Proc. of IEEE International Symposium on Information Theory (ISIT2010).}
\title{Capacity of a Class of Broadcast Relay Channels}

 \author{
\IEEEauthorblockN{Arash Behboodi and Pablo Piantanida} \\
\IEEEauthorblockA{Department of Telecommunications, SUPELEC\\
91192 Gif-sur-Yvette, France\\
Email: \{arash.behboodi,pablo.piantanida\}@supelec.fr
}
}

\maketitle

\begin{abstract}
Consider the broadcast relay channel (BRC) which consists of a source sending information over a two user broadcast channel in  presence of two relay nodes that help the transmission to the destinations. Clearly, this network with five nodes  involves all the problems encountered in relay and broadcast channels. New inner bounds on the capacity region of this class of channels are derived. These results can be seen as a generalization and hence unification of previous work in this topic. Our bounds are based on the idea of recombination of message bits and various effective coding strategies for relay and broadcast channels. Capacity result is obtained for the semi-degraded BRC-CR, where one relay channel is degraded while the other one is reversely degraded. An inner and upper bound is also presented for the degraded BRC with common relay (BRC-CR), where both the relay and broadcast channel are degraded which is the capacity for the Gaussian case. Application of these results arise in the context of opportunistic cooperation of cellular networks.
\end{abstract}   

\IEEEpeerreviewmaketitle

\section{Introduction}
Cooperative networks have been of huge interest during recent years between researchers. Using the multiplicity of information in nodes, these networks can provide the increase in capacity and reliability using the appropriate strategy. The simplest of these networks is the relay channel. A fundamental contribution was made by Cover and El Gamal \cite{Cover1979}, where the main strategies of Decode-and-Forward (DF) and Compress-and-Forward (CF), and an upper bound were developed for this channel along with capacity theorems for special classes of relay channels. Based on these strategies,  further work has been recently done on cooperative networks from different aspects, including deterministic channels \cite{Aref1982}, multiple access relay, broadcast relay and multiple relays, fading relay channels, etc. (see \cite{Kramer2005, Liang2007B, Laneman2004, Liang2007A} and references therein). 
Similarly, extensive research has been done on broadcast channels due to their importance as a main part of scenarios like multicast, flat, multi-hop, ad hoc, and others. This channel consists of a source transmitting different messages to several destinations. The main coding strategies (e.g. superposition coding, Marton coding) were developed in \cite{elgamal-vandermeulen-1981, Marton1979,  coverbroadcast-1972, ElGamal1979}, and shown to be capacity archiving for various classes of channels (e.g. degraded, degraded message sets, less noisy, more capable, deterministic). 

A variety of interesting networks that combine relay and broadcast channels have been also studied in \cite{Kramer2005,Liang2007B}. Coding techniques specific to each of these scenarios are merged together to derive achievable rates for broadcast relay channels. These results are of great interest  because many similar configurations can be found in practical network scenarios where these techniques can be used to characterize the basic limits of those networks. Connection between this class of channels and simultaneous relay channels are reported in \cite{Behboodi2009}.

In this paper we investigate the capacity region of the broadcast relay channel (BRC), which was first introduced in \cite{Behboodi2009}. This channel consists of a source transmitting common and private information to two destinations via a broadcast channel, in presence of two relay nodes that help this transmission. We derive new inner bounds on the capacity region of this class of channels. These bounds include Marton region \cite{Marton1979}, improve the  region first established in \cite{Behboodi2009} and the region previously derived by Kramer \textit{et al.} in \cite{Kramer2005}.  These regions are shown to be tight for the special cases of semi-degraded and degraded Gaussian broadcast relay channels. The techniques involved to obtain these results are essentially backward decoding, Marton coding,  reconfiguration of message  bits, motivated by  \cite{Kramer2005,Marton1979,Behboodi2009, Nair2009}.  Section II states definitions along with main results while the proof outlines are given in Sections III and IV. 

\section{Problem Definitions and Main Results}


\subsection{Problem Definition}

The Broadcast Relay Channel (BRC) consists of a source sending information to two destinations in presence of two helping relays as shown in Fig. \ref{fig:II-1}. This channel is defined by its stochastic mapping $\big\{P:\mathscr{X} \ex \mathscr{X}_1  \ex \mathscr{X}_2  \longmapsto \mathscr{Y}_1  \ex \mathscr{Z}_1 \ex \mathscr{Y}_2  \ex \mathscr{Z}_2 \big\}$, where the channel input is denoted by $X\in \mathscr{X}$, the relay inputs by $(X_1,X_2)\in \mathscr{X}_1 \times \mathscr{X}_2 $, the channel outputs by $(Y_1,Y_2)\in \mathscr{Y}_1 \times \mathscr{Y}_2 $ and the relay outputs by $(Z_1,Z_2)\in \mathscr{Z}_1 \times \mathscr{Z}_2 $. For sake of clarity, we define the notions of achievability for common and private rates $(R_0,R_1,R_2)$ and capacity, which remain the same as for  BCs \cite{coverbroadcast-1972}, \cite{Behboodi2009}.


\begin{definition}[Code] \label{def-code}
A code for the BRC consists of an encoder mapping $\{ \varphi:\mc{W}_{0} \ex\mc{W}_{1} \ex\mc{W}_{2}  \longmapsto \mathscr{X}^n  \}$, two decoder mappings $\{ \psi_t:\mathscr{Y}_t^n \longmapsto \mc{W}_{0} \ex\mc{W}_{t} \}_{t=\{1,2\}}$ and a set of relay functions $\{ f_{t,i} :\mathscr{Z}_t^{i-1}  \longmapsto \mathscr{X}_{ti}  \}_{i=1}^n$, for finite sets of integers $\mc{W}_{t}=\big\{ 1,\dots, W_{t} \big\}$. The rates of such code are $n^{-1} \log W_{t}$ with maximum error probabilities    
\begin{equation*}
e_{\max,t}^{(n)}\doteq \max_{(w_0,w_t)\in\mc{W}_{0}\ex\mc{W}_{t}}\Pr \big\{  \psi_t (\mb{Y} _t ) \neq (w_0,w_t)  \big\}.\label{errorprob_def}   
\end{equation*}    
\end{definition}
\begin{definition}[Achievable rates and capacity] 
For every $0$ \\$< \epsilon, \gamma< 1$, a triple of positive real numbers $(R_0,R_1,R_2)$ is achievable if for every sufficiently large $n$ there exist $n$-length block code satisfying $e_{\max,t}^{(n)}\big( \varphi,\psi_t, \{f_{t,i}\}_{i=1}^n\big) \leq \epsilon$ for $t=\{1,2\}$ and the rates 
$n^{-1} \log W_{t} \geq R_t-\gamma$ for $t=\{0,1,2\}$. The set of all achievable rates is called the capacity region. 
\end{definition}  

\subsection{Inner Bounds on the Capacity Region}
\begin{theorem} An inner bound on the capacity region of the broadcast relay channel is given by      
\begin{align*}
\mathscr{R}_I \doteq \displaystyle{\bigcup\limits_{P\in\mathscr{P}}}  \Big\{(R_0\geq 0,&R_1\geq 0,R_2 \geq 0): \\ 
R_0+R_1 &\leq I_1-I(U_0,U_1;X_2\vert X_1,V_0), \\
R_0+R_2 &\leq I_2-I(U_0,U_2;X_1\vert X_2,V_0), \\
R_0+R_1+R_2 &\leq I_1+J_2-I(U_0,U_1;X_2\vert X_1,V_0)-I(U_1,X_1;U_2\vert X_2,U_0,V_0)-I_M \\
R_0+R_1+R_2 &\leq J_1+I_2-I(U_0,U_2;X_1\vert X_2,V_0)-I(U_1;U_2,X_2\vert X_1,U_0,V_0)-I_M \\
2R_0+R_1+R_2 &\leq I_1+I_2-I(U_0,U_1;X_2\vert X_1,V_0)-I(U_0,U_2;X_1\vert X_2,V_0) \\
&-I(U_1;U_2\vert X_1,X_2,U_0,V_0)-I_M    \Big\}
\label{eq:II-1}     
\end{align*}
where $(I_i,J_i,I_M)$ with $i=\{1,2\}$ are as follows
\begin{equation*}
\begin{array} {l}
I_i\doteq \min\big\{I(U_0,U_i;Z_i\vert V_0,X_i)+I(U_{i+2};Y_i\vert U_0,V_0,X_i,U_i),I(U_0,V_0,U_i,U_{i+2},X_i;Y_i)\big\},\\
J_i\doteq \min\big\{I(U_i;Z_i\vert U_0,V_0,X_i)+I(U_{i+2};Y_i\vert U_0,V_0,X_i,U_i),I(U_{i+2},U_i,X_i;Y_i\vert U_0,V_0)\big\},  \\
I_M=I(U_3;U_4\vert U_1,U_2,X_1,X_2,U_0,V_0),
\end{array}   
\end{equation*}
and the union is over all joint PDs $P_{U_0V_0U_1U_2U_3U_4X_1X_2X}\in \mathscr{P}=\big \{P_{U_0V_0U_1U_2U_3U_4X_1X_2X}= P_{U_3U_4X|U_1U_2}\,P_{U_1U_2|U_0X_1X_2}\, P_{U_0|X_1X_2V_0}\,P_{X_2|V_0}\, P_{X_1|V_0}\,P_{V_0}$ with \\
$(U_0,V_0,U_1,U_2,U_3,U_4) \minuso (X_1,X_2,X) \minuso (Y_1,Z_1,Y_2,Z_2) \big \}$.   \vspace{1mm}
\label{thm:1}        
\end{theorem}

\begin{figure}[tb]
\centering
\subfigure[BRC with two relays]{
	\includegraphics [scale=0.5] {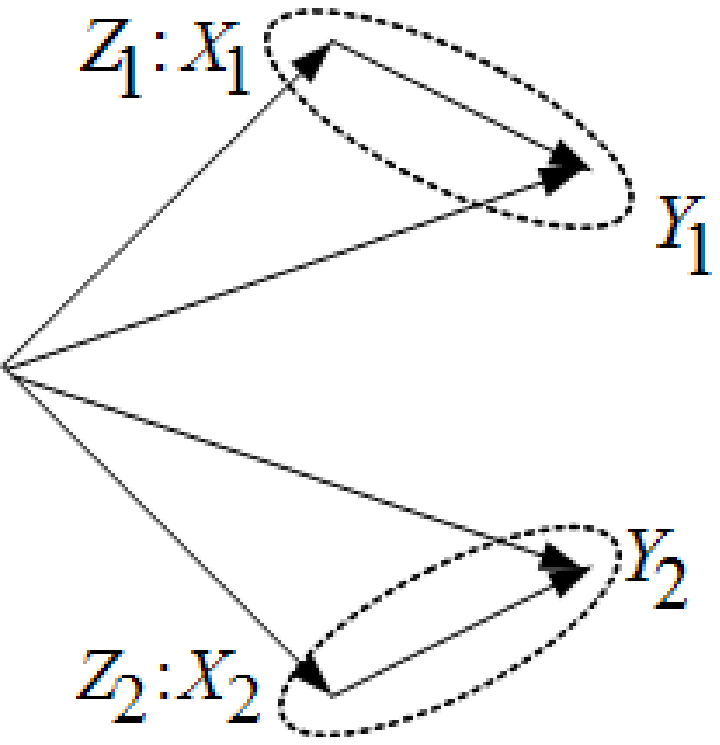}
\label{fig:II-1}
}
\subfigure[BRC with common relay]{
	\includegraphics [scale=0.5] {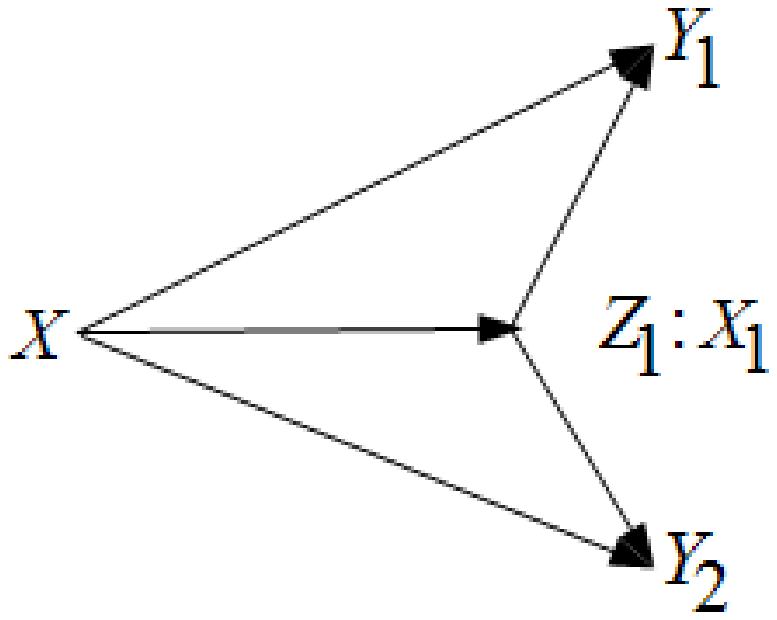}

\label{fig:II-2}
}
\vspace{-2mm}	\caption{Broadcast Relay Channel (BRC)}
\vspace{-4mm}
\end{figure}

\emph{Main ideas for coding:} The general coding idea in the theorem is depicted in Fig. \ref{fig:II-3}. The RV $V_0$ represents the common part of $(X_1,X_2)$ (the information sent at the relays) which is intended to help the common information encoded in $U_0$. Private information is sent in two steps, first  using the relay help through $(U_1,U_2)$ based on decode-and-forward (DF) strategy, and then by using the direct link between source and destinations to decode $(U_3,U_4)$. Marton coding is used to allow the correlation between the RVs denoted by arrows in Fig. \ref{fig:II-3}. We argue that both rates in theorem \ref{thm:1} coincide with the conventional rate based on partially DF  \cite{Cover1979}. 

\emph{Comparison to existent rate regions:} It is easy to verify that, by setting $(X_1,X_2,V_0,U_3,U_4)=\emptyset$, $Z_1=Y_1$ and $Z_2=Y_2$,  the rate region in theorem \ref{thm:1} includes Marton's region \cite{Marton1979}. Moreover, this region improves one derived for the BRC in \cite{Behboodi2009}, and for the single relay case depicted in Fig. \ref{fig:II-2}.
By choosing $X_2=U_2=\emptyset$ and $Z_2=(X,X_1)$, the rate region in theorem \ref{thm:1} can be shown to be a shaper inner bound than that previously found by Kramer \textit{et al.} in \cite{Kramer2005}, for example in the case of Gaussian degraded BRC-CR. The corresponding inner bound is given in the following corollary.         

\begin{corollary}
A sharper inner bound on the capacity region of the BRC-CR is given by 
\begin{equation*}
\begin{array} {l} 
\mathscr{R}_{II} \doteq \displaystyle{\bigcup\limits_{P_{V_0U_0U_1U_3U_4X_1X}\in\mathscr{P}}}  \Big\{(R_0\geq 0, R_1\geq 0,R_2 \geq 0):  \\ 
R_0+R_1 \leq \min\{I_1+I_{1p},I_3+I_{3p}\}+I(U_3;Y_1\vert U_1,U_0,X_1,V_0), \\
R_0+R_2 \leq I(U_0,V_0,U_4;Y_2)-I(U_0;X_1\vert V_0),  \\
R_0+R_1+R_2 \leq \min\{I_2,I_3\}+I_{3p}+I(U_3;Y_1\vert U_1,U_0,X_1,V_0)\\
\,\,\,\,\,\,\,\,\,\,\,\,\,+I(U_4;Y_4\vert U_0,V_0)-I(U_0;X_1\vert V_0)-I_M, \\
R_0+R_1+R_2 \leq \min\{I_2,I_1\}+I_{1p}+I(U_3;Y_1\vert U_1,U_0,X_1,V_0)\\
\,\,\,\,\,\,\,\,\,\,\,\,\,+I(U_4;Y_4\vert U_0,V_0)-I(U_0;X_1\vert V_0)-I_M, \\
2R_0+R_1+R_2  \leq I(U_3;Y_1\vert U_1,U_0,X_1,V_0)\\
\,\,\,\,\,\,\,\,\,\,\,\,\,+I(U_4;Y_4\vert U_0,V_0)+I_2+\min\{I_1+I_{1p},I_3+I_{3p}\}\\
\,\,\,\,\,\,\,\,\,\,\,\,\,-I(U_0;X_1\vert V_0)-I_M \Big\}
\label{eq:II-3}
\end{array}
\end{equation*}
where  $\mathscr{P}$ is the set of all joint PDs  $P_{V_0U_0U_1U_3U_4X_1X}$ satisfying that $(V_0,U_0,U_1,U_3,U_4) \minuso (X_1,X) \minuso (Y_1,Z_1,Y_2) \big \}$, and $I_1=I(U_0,V_0;Y_1)$, $I_2=I(U_0,V_0;Y_2)$, $I_3=I(U_0;Z_1|X_1,V_0)$, $I_{1p}=I(U_1X_1;Y_1|U_0,V_0)$, $I_{3p}=I(U_1;Z_1|U_0,V_0,X_1)$ and $I_M=I(U_3;U_4\vert X_1,U_1,U_0,V_0)$.   
\label{thm:2}
\end{corollary}

\subsection{Capacity Region of Degraded BRCs with Common Relay}

\begin{definition}[degraded BRC] \label{def-degraded}
A broadcast relay channel with common relay (BRC-CR) (as is shown in Fig. \ref{fig:II-2}) is said to be (or semi) degraded if the stochastic mapping $\big\{W:\mathscr{X} \ex \mathscr{X}_1  \ex \mathscr{X}_2  \longmapsto \mathscr{Y}_1  \ex \mathscr{Z}_1 \ex \mathscr{Y}_2  \ex \mathscr{Z}_2 \big\}$ satisfies one of the following Markov chains:
\begin{description}
\item[(I)] $X \minuso (X_1,Z_1) \minuso (Y_1,Y_2)$  and $(X,X_1) \minuso Y_1 \minuso Y_2$,
\item [(II)] $X \minuso (X_1,Z_1) \minuso Y_1$  and $X \minuso (Y_2,X_1) \minuso Z_1$,
\end{description}
where conditions (I) and (II) are referred to as degraded and semi-degraded BRC-CR, respectively.  
\end{definition}

\begin{theorem}
The upper bound on the capacity region of the degraded BRC-CR is given by the following rate region
\begin{align*} 
\mathscr{C}_{I}  \doteq \displaystyle{\bigcup\limits_{P_{UX_1X}\in\mathscr{P}}} \Big\{(R_0\geq 0 &,R_1\geq 0): \\
R_0  \leq &I(U;Y_2),\\ 
R_1  \leq &\min\big\{I(X;Z_1\vert X_1,U),I(X,X_1;Y_1\vert U)\big\}  \\
R_0+R_1 \leq &\min\big\{I(X;Z_1\vert X_1),I(X,X_1;Y_1)\big\}, \Big\}, 
\end{align*}
where $\mathscr{P}$ is the set of all joint PDs $P_{UX_1X}$ satisfying that $(V,U) \minuso (X_1,X) \minuso (Y_1,Z_1,Y_2)$. 
\label{thm:3}
\end{theorem}

\begin{conjecture}
The capacity region of the degraded BRC-CR is given by the following rate region
\begin{equation*}
\begin{array} {l}
R_0 \leq I(U,V;Y_2),\\ 
R_0+R_1 \leq \min\big\{I(X;Z_1\vert V,X_1),I(X,X_1;Y_1)\big\},  \\
R_0+R_1 \leq \min\big\{I(X;Z_1\vert X_1,V,U),I(X,X_1;Y_1\vert U,V)\big\}+I(V,U;Y_2) 
\end{array}
\end{equation*}
where $\mathscr{P}$ is the set of all joint PDs $P_{VUX_1X}$ satisfying that $(V,U) \minuso (X_1,X) \minuso (Y_1,Z_1,Y_2)$. 
\label{conj:1}
\end{conjecture}
The achievability part of this conjecture can be proved by choosing $U_3=U_4=\emptyset$ and by choosing $V_0=bU_0+(1-b)X_1$,where $b$ is a Bernoulli random variable with the parameter $p$ in the theorem \ref{thm:2}. The upper bound proof will be discussed while proving the theorem \ref{thm:3} later. The only difference between these two bounds are the additional conditions. Interestingly the bound presented in the conjecture and the upper bound in the theorem \ref{thm:3} happen to coincide for the case of Gaussian channel as presented in \cite{Bhaskaran2008} obtained via a different approach. With the definitions $Y_1=X+X_1+\mathpzc{N}_1$, $Y_2=X+X_1+\mathpzc{N}_2$, $Z_1=X+\tilde{\mathpzc{N}}_1$, with the source and the relay power constraint $P,P_1$ the following theorem holds similar to \cite{Bhaskaran2008}.
\begin{theorem}
The capacity region of the degraded Gaussian BRC-CR is
\begin{align*} 
R_0 & \leq C\left(\frac{\alpha(P+P_1+2\sqrt{\overline\beta PP_1})}{\overline\alpha(P+P_1+2\sqrt{\overline\beta PP_1})+{N}_2}\right),\\ 
R_1 & \leq  C\left(\frac{\overline\alpha(P+P_1+2\sqrt{\overline\beta PP_1})}{{N}_1}\right), 
R_1 \leq C\left(\frac{\beta\gamma P}{\tilde{N}_1}\right) \\
R_0+R_1 & \leq  C\left(\frac{\beta P}{\tilde{N}_1}\right) 
\end{align*}
where $0\leq\beta,\alpha,\gamma\leq1$.
\label{thm:3-1}
\end{theorem}
The capacity region of semi-degraded BRC-CR is stated in the following theorem. 
\begin{theorem}
The capacity region of the semi-degraded BRC-CR is given by the following rate region
\begin{align*} 
\mathscr{C}_{II} & \doteq \displaystyle{\bigcup\limits_{P_{U_0X_1X}\in\mathscr{P}}} \Big\{(R_0\geq 0 ,R_1\geq 0): \\
R_0 & \leq \min\{I(U_0,X_1;Y_1),I(Z_1;U_0\vert X_1)\}\\ 
R_0+R_1 & \leq  \min\{I(U_0,X_1;Y_1),I(Z_1;U_0\vert X_1)\} +I(X;Y_2\vert X_1,U_0) \Big\}, 
\end{align*}
where $\mathscr{P}$ is the set of all joint PDs $P_{U_0X_1X}$ satisfying that $U_0 \minuso (X_1,X) \minuso (Y_1,Z_1,Y_2)$. 
\label{thm:4}
\end{theorem}
%
It easy to show that  the rate region stated in theorem \ref{thm:4} directly follows from  that of theorem \ref{thm:1}, by setting  $X_1=X_2=V_0$, $Z_1=Z_2$, $U_1=U_2=U_3=\phi$ and $U_4=X$. 

\section{Sketch of Proof of Theorem \ref{thm:1}} \label{proof}

 First, split the private information $W_b$ into non-negative indices $(S_{0b},S_b,S_{b+2})$ with $b\in\{1,2\}$. Then merge the common information $W_0$ with a part of private information $(S_{01},S_{02})$ into a single message. Thus we have that $R_b= S_{b+2}+S_{b}+S_{0b}$. 
\textit{Code Generation:}
\begin{enumerate}[(i)]
	\item Generate $2^{nT_0} $ i.i.d. sequences $\underline{v}_0$ each from  $P_{V_0}(\underline{v}_0)=\prod_{j=1}^n p_{V_0}(v_{0j})$ indexed as $\underline{v}_0(r_0)$ with $r_0\in \left[1, 2^{nT_0} \right]$.
	\item 
	For each $\underline{v}_0(r_0)$, generate $ 2^{nT_0} $ i.i.d. sequences   $\underline{u}_0$ each from $P_{U_0| V_0}(\underline{u}_0\vert \underline{v}_0(r_0))=\prod_{j=1}^np_{U_0\vert V_0}(u_{0j}\vert v_{0j}(r_0))$. Index them as $\underline{u}_0(r_0,t_0)$ with $t_0\in \left[1,2^{nT_0}\right]$. 
	\item
	For $b\in\{1,2\}$ and each $\underline{v}_0(r_0)$, generate $2^{nT_b}$ i.i.d. sequences $\underline{x}_b$ each from $P_{X_b| V_0}(\underline{x}_b\vert \underline{v}_0(r_0))=\prod_{j=1}^np_{X_b\vert V_0}(x_{bj}\vert v_{0j}(r_0))$.
	Index them as $\underline{x}_b(r_0,r_b)$ with $r_b\in \left[1,2^{nT_b}\right]$.  
\item
Partition the set $\big\{1,\ldots,2^{nT_0}\big\}$ into $2^{n(R_0+S_{01}+S_{02})}$ cells (similarly to \cite{Marton1979}) and label them as $S_{w_0,s_{01},s_{02}}$. In each cell there are $2^{n(T_0-R_0-S_{01}-S_{02})}$ elements. 	
\item
For each $\underline{v}_0(r_0)$, the encoder searches for an index $t_{0}$ at the cell $S_{w_{0},s_{01},s_{02}}$ such that $\underline{u}_0\big(r_{0},t_{0}\big)$ is jointly typical with $\big(\underline{x}_{1}(r_{0}, r_{1}),\underline{x}_{2}(r_{0}, r_{2}),\underline{v}_0(r_{0})\big)$. The successful of this step requires that \cite{Marton1979}.
\begin{equation}
T_0-R_0-S_{01}-S_{02}\geq I(U_0;X_1,X_2\vert V_0). 
\label{eq:III-1}
\end{equation}
\item
For each $b \in \{1,2\}$ and every typical pair $\big(\underline{u}_0(r_{0},t_{0}),$ $\underline{x}_b(r_{0},r_{b}) \big)$ chosen in the bin $(w_{0},s_{01},s_{02})$, generate $2^{nT_b}$ i.i.d. sequences $\underline{u}_{b}$ each from  $P_{U_b| X_b,U_0}\big (\underline{u}_{b}\vert \underline{u}_0(r_{0},t_{0}),\underline{x}_b(r_{0},r_{b}), \underline{v}_0(r_{0}) \big)= \prod_{j=1}^np_{U_b\vert UX_b V}(u_{bj}\vert u_{0j}(r_{0},t_{0}),x_{bj}(r_{0},r_{b}),v_{0j}(r_{0})) $. Index them as $\underline{u}_b(r_{0},t_{0},r_{b},t_{b})$ with $t_b\in \left[1,2^{nT_b}\right]$.  
\item
For $b\in\{1,2\}$, partition the set $\big\{1,\ldots,2^{nT_b}\big\}$ into $2^{nS_b}$ cells and label them as $S_{s_b}$. In each cell there are $2^{n(T_b-S_b)}$ elements.  
\item \label{FirstMarton}
For each $b\in\{1,2\}$ and every cell $S_{s_{b}}$, define the set $\mathscr{L}_b$ to be the set of all sequences  $\underline{u}_b\big(r_{0},t_{0},r_{b},t_{b}\big)$ for $t_{b}\in{S}_{s_{b}}$ that are jointly typical with $\big( \underline{x}_{\overline{b}}(r_{0}, r_{\overline{b}}),$ $\underline{v}_0(r_{0}), \underline{u}_0(r_{0},t_{0}),\underline{x}_b(r_{0},r_{b})\big)$, where $\overline{b}=\{1,2\}\setminus \{b\}$. In order to creat $\mathscr{L}_b$, we look for the $\underline{u}_b$-index inside the cell $S_{s_{b}}$ and find $\underline{u}_b$ such that it belongs to the set of $\epsilon$-typical $n$-sequences $\textsl{A}^n_\epsilon(V_0U_0X_1X_2U_b)$.
\item  \label{SecondMarton}
Then search for a pair $(\underline{u}_1\in\mathscr{L}_1,\underline{u}_2\in\mathscr{L}_2)$ such that $\big(\underline{u}_1(r_{0},t_{0},r_{1},t_{1}),\underline{u}_2(r_{0},t_{0},r_{2},t_{2})\big) $ are jointly typical given the RVs $\big(\underline{v}_0(r_{0}), \underline{x}_2(r_{0},r_{2}),\underline{x}_1(r_{0},r_{1}),\underline{u}_0(r_{0},t_{0})\big)$. The success of coding steps (\ref{FirstMarton}) and (\ref{SecondMarton}) requires 
\begin{align}
T_b-S_b & \geq I(U_b;X_{\overline{b}}\vert X_b,U_0,V_0), \nonumber\\
T_1+T_2-S_1-S_2 & \geq I(U_1;X_2\vert X_1,U_0,V_0)  \\ 
+ I(U_2;X_1\vert X_2,U_0,V_0) & +I(U_2;U_1\vert X_1,X_2,U_0,V_0).\nonumber \label{eq:III-2}
\end{align}

\begin{figure}[t]
\centering
	\includegraphics [scale=0.3] {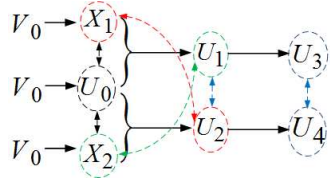}
\caption{Diagram of auxiliary random variables}    \vspace{-4mm}
\label{fig:II-3}
\end{figure}
The first inequality for $b\in \{1,2\}$ guarantees the existence of non-empty sets $(\mathscr{L}_1,\mathscr{L}_2)$ and the last one is for the step (\ref{FirstMarton}).
\item
$(\underline{u}_3,\underline{u}_4)$: For each $b\in\{1,2\}$ and every typical pair of sequences $\big( \underline{u}_1(r_{0},t_{0},r_{1},t_{1}),$ $\underline{u}_2(r_{0},t_{0},r_{2},t_{2}) \big)$ chosen in the bin $(s_{1},s_{2})$, generate $2^{nT_{b+2}}$ i.i.d. sequences $\underline{u}_{b+2}$ each i.i.d. from $P_{U_{b+2}| U_b }(\underline{u}_{b+2}\vert \underline{u}_b(r_{0},t_{0},r_{b},t_{b}))= \prod_{j=1}^np_{U_{b+2}\vert U_b}(u_{(b+2)j}\vert u_{bj}(r_{0},t_{0},r_{b},t_{b}))$. Index them as $\underline{u}_{b+2}(r_{0},t_{0},r_{b},t_{b},t_{b+2})$ with $t_{b+2}\in \left[1,2^{nT_{b+2}}\right]$.
\item 
For $b\in\{1,2\}$, partition the set $\big\{1,\ldots,2^{nT_{b+2}}\big\}$ into $2^{nS_{b+2}}$ cells and label them as $S_{s_{b+2}}$. In each cell there are $2^{n(T_{b+2}-S_{b+2})}$ elements. 	
\item
The encoder searches for index $t_{3}\in S_{s_{3}}$ and $t_{4}\in S_{s_{4}}$, such that $\underline{u}_3\big(r_{0},t_{0},r_{1},t_{1},t_3\big)$ and $\underline{u}_4\big(r_{0},t_{0},r_{2},t_{2},t_4\big)$ are jointly typical given each chosen typical pair of $\underline{u}_1(r_{0},t_{0},r_{1},t_{1})$ and $\underline{u}_2(r_{0},t_{0},r_{2},t_{2})$.
The success of this encoding step requires
\begin{equation}
\!\!\!\! \!\!\!\! T_3+T_4-S_{3}-S_{4}\geq I(U_3;U_4\vert U_1,U_2,X_1,X_2,U_0,V_0).  
\label{eq:III-3}
\end{equation}
\end{enumerate}
\textit{Encoding Part:} The transmission is done in $B+1$ block. The encoding in block $i$ is as follows:
\begin{enumerate}[(i)]
\item
First, reorganize the current message $(w_{0i},w_{1i},w_{2i})$ into $(w_{0i},s_{01i},s_{02i},s_{1i},s_{2i},s_{3i},s_{4i})$.
\item
Then for each $b\in\{1,2\}$, relay $b$ already knows about the indices $(t_{0(i-1)},t_{b(i-1)})$, so it sends $\underline{x}_b\big(t_{0(i-1)},t_{b(i-1)}\big)$.
\item
Once the encoder found $(t_{0i},t_{1i},t_{2i},t_{3i},t_{4i})$ (based on the code generation) corresponding to $(w_{0i},s_{01i},s_{02i},s_{1i},s_{2i},s_{3i},s_{4i})$ the source transmits $\underline{x}(r_{0(i-1)},t_{0i},r_{1(i-1)},r_{2(i-1)},t_{1i},t_{2i},t_{3i},t_{4i})$ which is randomly drawn from $P_{U_1U_2U_3U_4}(\underline{u}_1,\underline{u}_2,\underline{u}_3,\underline{u}_4)$.
\end{enumerate}
\textit{Decoding Part:} 
\begin{enumerate}[(i)]
\item
First for $b\in\{1,2\}$, the relay $b$ after receiving $z_{bi}$ tries to decode $(t_{0i},t_{bi})$. The relay is aware of $(V_0,X_b)$ because it is supposed to know about $(t_{0(i-1)},t_{b(i-1)})$. The relay $b$ declares that the pair $(t_{0i},t_{bi})$ is sent if the following conditions are simultaneously satisfied:
\begin{enumerate} 
\item
$\underline{u}_0(t_{0(i-1)},t_{0i})$ is jointly typical with $\big(z_{bi}$, $\underline{v}_0(t_{0(i-1)}),$ $\underline{x}_b (t_{0(i-1)},t_{b(i-1)}) \big)$. 
\item
$\underline{u}_b(t_{0(i-1)},t_{0i},t_{b(i-1)},t_{bi})$ is jointly typical with $\big(z_{bi}$, $\underline{v}_0(t_{0(i-1)}),$ $\underline{x}_b(t_{0(i-1)},t_{b(i-1)}) \big)$. 
\end{enumerate} 
Notice that $\underline{u}_0$ has been generated independent of $\underline{x}_b$ and hence $\underline{x}_b$ does not appear 
in the given part of mutual information. This is an important issue that may increase the region. Constraints for reliable decoding are:
\begin{align}
T_b & < I(U_b;Z_b\vert U_0,V_0,X_b), \label{eq:III-4A} \\
\!\!\!\! T_b+T_0 & < I(U_b;Z_b\vert U_0,V_0,X_b)+I(U_0;Z_b,X_b\vert V_0). \!\!\!\! \label{eq:III-4B}
\end{align}
\begin{remark}
The intuition behind expressions \eqref{eq:III-4A} and \eqref{eq:III-4B} is as follows. Since the relay knows $\underline{x}_{b(i-1)}$ we are indeed decreasing the cardinality of the set of possible $\underline{u}_0$,  which without additional knowledge is $2^{nT_0}$. The new set of possible $(\underline{u}_0$, $\mathscr{L}_{X_b})$ can be defined as all $\underline{u}_0$ jointly typical with $\underline{x}_{b(i-1)}$. It can be shown \cite{elgamal-vandermeulen-1981} that $\esp[\left\|\mathscr{L}_{X_b}\right\|]=2^{n [T_0-I(U_0;X_b\vert V_0) ] }$, which proves our claim on the reduction of cardinality.  
One can see that after simplification \eqref{eq:III-4B} using \eqref{eq:III-1}, $I(U_0;Z_b,X_b\vert V_0)$ is removed and the final bound reduces to $I(U_0,U_b;Z_b\vert V_0,X_b)$.
\end{remark}
\item 
For each $b\in\{1,2\}$ destination $b$, after receiving $y_{b(i+1)}$, tries to decode  the relay-forwarded information $(t_{0i},t_{bi})$, knowing $(t_{0(i+1)},t_{b(i+1)})$. It also tries to decode the direct information $t_{(b+2)(i+1)}$. Backward decoding is used to  decode index $(t_{0i},t_{bi})$. The decoder declares that $(t_{0i},t_{bi},t_{(b+2)(i+1)})$ is sent if the following constraints are simultaneously satisfied:
\begin{enumerate} 
\item 
$\big(\underline{v}_0(t_{0i}),\underline{u}_0(t_{0i},t_{0(i+1)}),y_{b(i+1)}\big)$  are jointly typical,
\item \label{BackwardDecodingI}
$\big(\underline{x}_b(t_{0(i)},t_{b(i)}), \underline{v}_0(t_{0i}), \underline{u}_0(t_{0i},t_{0(i+1)})  \big)$  and $y_{b(i+1)}$ are jointly typical,
\item
$ \big( \underline{u}_b(t_{0i},t_{0(i+1)},t_{bi},t_{b(i+1)}),\underline{u}_{b+2}(t_{0i},t_{0(i+1)},t_{bi},t_{(b+2)b(i+1)},t_{b(i+1)})  \big)$ and  $\big( y_{b(i+1)}, \underline{v}_0(t_{0i}),\underline{u}_0(t_{0i},t_{0(i+1)}), \underline{x}_b\big(t_{0(i)},$ $ t_{b(i)}\big)\big)$ are jointly typical. 
\end{enumerate}
Notice that in the decoding step (\ref{BackwardDecodingI}) the destination knows about $t_{0(i+1)}$, which has been chosen such that $(\underline{u}_0,\underline{x}_b)$ are jointly typical and this information contributes to decrease the cardinality of all possible $\underline{x}_b$ (similarly to what happened in decoding at the relay). Hence $U_0$ in step (\ref{BackwardDecodingI}) does not  appear in the given part of mutual information. From this we have that the main constraints for successful decoding are as follows:
\begin{align}
T_{b+2} & < I(U_{b+2};Y_b\vert U_0,V_0,X_b,U_b),\label{eq:III-6A}\\
T_{b+2}+T_b & < I(U_{b+2},U_b,X_b;Y_b\vert U_0,V_0), \label{eq:III-6B} \\
T_{b+2}+T_b+T_0 & < I(V_0,U_0;Y_b)+I(X_b;Y_b,U_0\vert V_0) +I(U_{b+2},U_b;Y_b\vert U_0,V_0,X_b). \label{eq:III-6C}
\end{align}
Observe that $U_0$ increases the bound in \eqref{eq:III-6B}. Similarly using \eqref{eq:III-1} and after removing the common term $I(U_0;X_b\vert V_0)$, one can simplify the bound in \eqref{eq:III-6C}, to $I(U_{b+2},U_b,X_b,V_0,U_0;Y_b)$.
\item Theorem \ref{thm:1} follows by applying Fourier-Motzkin elimination to  
\eqref{eq:III-1}-\eqref{eq:III-6C} and using the non-negativity of the rates. 
\end{enumerate}

\section{Sketch of Proof of Theorems \ref{thm:3} and \ref{thm:4}}

\subsection{Proof of the Upper Bound} 
We now prove the upper bound in the theorem \ref{thm:3} is the capacity of degraded BRCs. First, notice that the second bound is the capacity of a degraded relay channel shown in \cite{Cover1979}. 
Regarding the fact that user $1$ is decoding all the information, the bound can be reached using the same method. So we focus on the reminder bounds. For any code $(n,\mc{M}_0,\mc{M}_1,e_{\max}^{(n)})$ (i.e. $(R_0,R_1)$), we want to show that if the error probability goes to zero then, the rates satisfy the conditions in theorem \ref{thm:3}. From Fano's inequality we have that  
\begin{align*}
H(W_0|{Y}_0) & \leq e_{\max}^{(n)}nR_0+1\stackrel{\Delta}{=}n\epsilon_0, \\
H(W_0,W_1|{Y}_1) & \leq e_{\max}^{(n)}n(R_0+R_1)+1\stackrel{\Delta}{=}n\epsilon_1,  
\end{align*}
and           \vspace{-4mm}
 \begin{align*}
nR_0  & \leq I(W_0;Y_2)+n\epsilon_0,\\
n(R_0+R_1) &  \leq I(W_0;Y_2)+I(W_1;Y_1\vert W_0)+n\epsilon_0+n\epsilon_1.
 \end{align*}
By setting $U_i=(Y_2^{i-1},W_0)$, then it can be shown that
\begin{equation*}
\begin{array} {l} 
I(W_1;Y_1\vert W_0)=
\displaystyle\sum_{i=1}^nI(W_1;Y_{1i}\vert Y_{1}^{i-1},W_0)=\\
\displaystyle\sum_{i=1}^nH(Y_{1i}\vert Y_{1}^{i-1},W_0)-H(Y_{1i}\vert Y_{1}^{i-1},W_0,W_1)\stackrel{(a)}{\leq} \\
\displaystyle\sum_{i=1}^nH(Y_{1i}\vert Y_{2}^{i-1},W_0)-H(Y_{1i}\vert X_{i},X_{1i},Y_{1}^{i-1},W_0,W_1) \\
\displaystyle \stackrel{(b)}{=}\sum_{i=1}^nH(Y_{1i}\vert Y_2^{i-1},W_0)-H(Y_{1i}\vert X_{i},X_{1i}) \stackrel{(c)}{\leq} \\
\displaystyle\sum_{i=1}^nI(X_{i},X_{1i};Y_{1i}\vert Y_2^{i-1},W_0)= \sum_{i=1}^nI(X_{i},X_{1i};Y_{1i}\vert U_i), \\
\end{array}
\label{eq:IV-3-2} \vspace{-2mm}
\end{equation*}
where $(a)$ results from the degradedness between $Y_{1}$ and $Y_2$, where (b) and (c) require Markov chain $Y_{1i}$ and $(X_{i},X_{1i})$.
\begin{equation*}
\begin{array} {l} 
I(W_1;Y_1\vert W_0)\leq I(W_1;Y_1,Z_1\vert W_0)=\\
\displaystyle\sum_{i=1}^nH(W_1\vert Y_{1}^{i-1},Z_{1}^{i-1},W_0)-H(W_1\vert Y_{1}^{i},Z_{1}^{i},W_0)\stackrel{(d)}{\leq} \\
\displaystyle\sum_{i=1}^nH(W_1 \vert Z_{1}^{i-1},X_{1i},W_0)-H(W_1\vert X_{1i},Z_{1}^{i},W_0) = \\
\displaystyle\sum_{i=1}^nH(Z_{1i}\vert Z_{1}^{i-1},X_{1i},W_0)-H(Z_{1i}\vert X_{1i},Z_{1}^{i-1},W_0,W_1) \leq \\
\end{array}
\end{equation*}
\begin{equation*}
\begin{array} {l} 
\displaystyle\sum_{i=1}^nH(Z_{1i}\vert Z_{1}^{i-1},X_{1i},W_0)-H(Z_{1i}\vert X_{i},X_{1i},Z_{1}^{i-1},W_0,W_1)\stackrel{(e)}{\leq} \\
 \displaystyle \sum_{i=1}^nH(Z_{1i}\vert Y_2^{i-1},X_{1i},W_0)-H(Z_{1i}\vert X_{i},X_{1i}) \stackrel{(f)}{=} 
  \end{array}
\label{eq:IV-4A} \vspace{-2mm}
\end{equation*}
 \begin{equation*} 
\begin{array} {l} 
\displaystyle\sum_{i=1}^nH(Z_{1i}\vert Y_2^{i-1},X_{1i},W_0)-H(Z_{1i}\vert X_{i},X_{1i},Y_2^{i-1},W_0) = \\
\displaystyle\sum_{i=1}^nI(X_{i};Z_{1i}\vert X_{1i},Y_2^{i-1},W_0)= \sum_{i=1}^nI(X_{i};Z_{1i}\vert X_{1i},U_i). 
\end{array}
\label{eq:IV-4B} \vspace{-2mm}
\end{equation*}
Based on the definition $X_{1i}$ can be obtained via $Z_1^{i-1}$, so given $Z_1^{i-1}$ one can have $X_1^{i-1}$, and then with $Z_1^{i-1},X_1^{i-1}$ and using Markovity between $(Z_{1},X_1)$ and $(Y_1,Y_2)$, one can say that $(Y_1^{i-1},Y_2^{i-1})$ is also available given $Z_1^{i-1}$. Step (d) and (e) result from this fact. Markovity of $Z_{1i}$ and $(X_{i},X_{1i})$ has been used for (e) and (f). For the first inequality, we have
\begin{equation}
\begin{array} {l} 
I(W_0;Y_2)=\displaystyle\sum_{i=1}^nI(W_0;Y_{2i}\vert Y_{2}^{i-1}) \leq  \sum_{i=1}^nI(U_{i};Y_{2i})
\end{array}
\label{eq:IV-5} \vspace{-2mm}
\end{equation}
Finally the bound can be proved using an independent time sharing RV $Q$ as is done in \cite{Cover1979}.
The upper bound for the semi-degraded BRC can be proved with the exact same technique. However regarding the space limit it is not presented here. On the other hand the upper bound presented in the conjecture is essentially same is this region by defining $V_{i}=Y_{2}^{i-1}$. One can see that $V_{i}$ can be inserted with $U_{i}$ everywhere. 
\subsection{Proof of the theorem \ref{thm:3-1}} 
The achievability of the rate can be established using the inner bound presented at the conjecture and in the same way as \cite{Bhaskaran2008}, so is not presented here. But the upper bound is calculated using the theorem \ref{thm:3}. We start by: 
\begin{equation*}
\begin{array} {l} 
I(U;Y_2)=h(Y_2)-h(Y_2|U)\\
h(Y_2)\leq \frac{n}{2}\log(2\pi e(N_2+P+P_1+2\sqrt{\overline{\beta}PP_1}))\\
h(\mathpzc{N}_2)\leq h(Y_2|U)\leq h(Y_2)
\end{array}
\end{equation*}
so there is $\alpha$ such that $h(Y_2|U)=\frac{n}{2}\log(2\pi e(N_2+\alpha(P+P_1+2\sqrt{\overline{\beta}PP_1})))$. While using the power entropy inequality we have:
\begin{equation*}
\begin{array} {l} 
e^{\frac{2}{n}h(Y_1|U)} \leq e^{\frac{2}{n}h(Y_2|U)}-e^{\frac{2}{n}h(\mathpzc{N}_2-\mathpzc{N}_1)}
\end{array}
\end{equation*}
and hence $h(Y_1|U)\leq \frac{n}{2}\log(2\pi e(N_1+\alpha(P+P_1+2\sqrt{\overline{\beta}PP_1})))$. On the other hand we have:
\begin{equation*}
\begin{array} {l} 
I(X,X_1;Y_1|U)=h(Y_1|U)-h(Y_1|X,X_1,U)\\
h(Y_1|X,X_1,U)=h(\mathpzc{N}_1)
\end{array}
\end{equation*}
Using the constraints for $h(Y_1|U),h(Y_2|U),h(Y_1|X,X_1,U),h(Y_2)$, the bounds are easily obtained. The calculation of $I(X;Z_1|X_1)$ is done like \cite{Cover1979} by bounding $h(Z_1|X_1)\leq \frac{n}{2}\log(2\pi e(\tilde{N}_2+\beta P))$. Finally we have:
\begin{equation*}
\begin{array} {l} 
I(X;Z_1|U,X_1)=h(Z_1|U,X_1)-h(Z_1|X,X_1)\\
h(\tilde{\mathpzc{N}}_1)\leq h(Z_1|U,X_1)\leq h(Z_1|X_1)\\
h(Z_1|X,X_1)=h(\tilde{\mathpzc{N}}_1)
\end{array}
\end{equation*}
Using the bound of $h(Z_1|X_1)$, it can be said that there is $\gamma$ such that $h(Z_1|X_1,U)= \frac{n}{2}\log(2\pi e(\tilde{N}_1+\beta\gamma P))$. After bounding this rate, the rest of the proof goes as usual. 

\section{Summary and Discussions}

A general achievable rate region has been developed by combining various well-known techniques including  recombination of message bits, Marton coding, backward decoding  and improved also by using novel techniques. The region includes Marton region for BCs, improves previous results by Kramer \textit{et al.} \cite{Kramer2005} and also existing rate regions  on this class of channels \cite{Behboodi2009}. Capacity is shown for the cases of semi-degraded and degraded Gaussian BRCs with common relays. On-going work includes capacity for other special cases as semi-deterministic relay channels and others.

\bibliographystyle{IEEEtran}
\bibliography{biblio}
\end{document}